\documentclass[preprint]{aastex}
\slugcomment{The Astronomical Journal, in press}
\shorttitle{Shapley Supercluster collapse}
\shortauthors{Reisenegger et al.}

\begin{document}

\title{The Shapley Supercluster. III. Collapse dynamics and mass of the central
concentration}

\author{Andreas Reisenegger\altaffilmark{1}, H. Quintana\altaffilmark{1}, 
Eleazar R. Carrasco\altaffilmark{2}, and Jer\'onimo Maze\altaffilmark{1}}

\altaffiltext{1}{Departamento de Astronom\'{\i}a y Astrof\'{\i}sica, Facultad de
F\'\i sica, Pontificia Universidad Cat\'olica de Chile, Casilla 306, 
Santiago 22, Chile}
\altaffiltext{2}{Instituto Astron\^omico e Geof\'{\i}sico, Universidade de S\~ao 
Paulo, Caixa Postal 3386, 01060-970, S\~ao Paulo, Brazil}
\email{areisene@astro.puc.cl}

\begin{abstract}
We present the first application of a spherical collapse model to
a supercluster of galaxies. Positions and redshifts of $\sim 3000$ 
galaxies in the Shapley Supercluster (SSC) are used to define velocity caustics
that limit the gravitationally collapsing structure in its central part. 
This is found to extend at least to $8 h^{-1}$ Mpc of the central cluster, 
A 3558, enclosing 11 ACO clusters. Infall velocities reach $\sim 2000$ km s$^{-1}$. 
Dynamical models of the
collapsing region are used to estimate its mass profile. An upper bound
on the mass, based on a pure spherical infall model, gives 
$M(<8 h^{-1}{\rm Mpc})\,^{<}_{\sim}\,1.3\times 10^{16}h^{-1}M_\odot$ for an
Einstein-de Sitter (critical) Universe and 
$M(<8 h^{-1}{\rm Mpc})\,^{<}_{\sim}\,8.5\times 10^{15}h^{-1}M_\odot$ 
for an empty Universe.
The model of Diaferio \& Geller (1997), based on estimating the
escape velocity, gives a significantly lower value, 
$M(<8 h^{-1}{\rm Mpc})\approx 2.1\times 10^{15}h^{-1}M_\odot$, very similar 
to the mass found around the Coma cluster by the same method (Geller et al. 
1999), and comparable to or slightly lower than the dynamical mass 
in the virialized regions of clusters enclosed in the same region of the SSC. 
In both models, the overdensity in this region is
substantial, but far from the value required to account for the peculiar
motion of the Local Group with respect to the cosmic microwave background.
\end{abstract}


\newpage

\section{Introduction}

This is the third in a series of papers analyzing the structure and physical
parameters of the Shapley Supercluster (SSC), based on galaxy redshifts. 
The first (Quintana et al. 1995; hereafter {\it Paper I}) presents and 
gives an initial analysis of the results of 
spectroscopic observations of the central region. The second (Quintana, 
Carrasco, \& Reisenegger 2000, hereafter {\it Paper II}) presents a much extended 
sample of galaxy redshifts and gives a qualitative discussion of the 
SSC's morphology. Here, we use dynamical collapse models
applied to this sample, in order to obtain the mass of the central
region of the SSC. An upcoming paper (Carrasco, Quintana, 
\& Reisenegger 2000, hereafter {\it Paper IV}) will analyze the 
individual clusters of galaxies contained in the 
sample, to obtain their physical parameters (velocity dispersion, size,
mass), search for substructures within the clusters, and determine
the total mass contained within the virialized regions of clusters in
the whole Shapley area. 

The Shapley concentration (Shapley 1930) is the richest supercluster in the 
local Universe (Zucca et al. 1993, Einasto et al. 1997, but see also
Batuski et al. 1999). This makes its study important for three main reasons. 
First, its high density of mass 
and of clusters of galaxies provides an extreme environment in which to
study galaxy and cluster evolution. Second, its existence and the fact
that it is the richest supercluster in a given volume constrain theories of
structure formation, and particularly the cosmological parameters and
power spectrum in the standard model of hierarchical structure formation
by gravitational instability (e.g., Ettori, Fabian, \& White 1997; 
Bardelli et al. 2000). 
Finally, it is located near the apex of the motion of the Local 
Group with respect to the cosmic microwave background. Thus it is intriguing
whether the SSC's gravitational pull may contribute significantly to 
this motion, although most mass estimates (e.g., Raychaudhury 1989; 
Raychaudhury et al. 1991; Paper I; Ettori et al. 1997; Bardelli et al. 2000) 
make a contribution beyond a 10 \% level very unlikely. 

Galaxy counts in redshift space (Bardelli et al. 2000) 
suggest that most of the supercluster has a density several times the
cosmic average, while the two complexes within $\sim 5 h^{-1}$ Mpc of 
clusters A 3558 and A 3528 have overdensities $\sim 50$
and $\sim 20$, respectively.
These regions are therefore far outside the ``linear regime'' of small
density perturbations, but still far from being virialized after full
gravitational collapse. The same conclusions are easily reached by even a
casual glance at the redshift structure presented in Paper II.
The density of these complexes indicates that they should be presently
collapsing (e.g., Bardelli et al. 2000), and in the present paper we
study this hypothesis for the main complex, around A 3558, where substantially
more data, with better areal coverage, are available (see Paper II).

It should be pointed out that within each of these complexes we expect very
large peculiar velocities, which dominate by far over their Hubble expansion.
In this case, redshift differences among objects within each complex can give
information about its dynamics, which we will analyze below, but essentially
no information about relative positions along the line of sight (except, 
perhaps, non-trivial information within a given dynamical model). While this
is generally acknowledged to be true within clusters of galaxies, where
galaxy motions have been randomized by the collapse, it is sometimes
overlooked on somewhat larger, but still nonlinear scales. For instance,
Ettori et al. (1997) calculate three-dimensional distances between clusters
in the Shapley region on the basis of their angular separations and redshifts,
and conclude that the group SC 1327-312 and the cluster A 3562, at projected 
distances $\sim 1$ and $3 h^{-1}$ Mpc from the central cluster A 3558, are 
between 5 and $10 h^{-1}$ Mpc from it in three-dimensional space, because of 
moderate differences in redshift. However, there is evidence for interactions 
between these clusters and groups (Venturi et al. 1999), suggesting true 
distances much closer to the projected distances. The discrepancy is 
naturally explained by quite modest peculiar velocities of several hundreds of 
km s$^{-1}$, easily caused by the large mass concentration.

In the present paper, we analyze the region around A 3558 in terms of an
idealized, spherical collapse model, which is used both in its pristine,
but undoubtedly oversimplified, original form (Reg\"os \& Geller 1989),
seen from a slightly different point of view, and in its less appealing,
but possibly more accurate, modern fine-tuning calibrated by simulations 
(Diaferio \& Geller 1997; Diaferio 1999). Section 2 explains the
models, argues for the presence of velocity caustics, and gives the 
equations relating the caustics position to the mass distribution. 
(Some mathematical remarks regarding this relation are given in the Appendix.)
In \S 3 we present the data, argue that velocity caustics are indeed present,
and explain how we locate their position quantitatively. Section 4 presents
and discusses our results, and \S 5 contains our main conclusions.

\section{The models}

\subsection{Pure spherical collapse}

In this approach, we consider a spherical structure in which matter
at any radius $r$ moves radially, with its acceleration determined by
the enclosed mass $M(r)$. At a given time $t_1$ of observation, the infall velocity 
$u(r)=-\dot r(r)$ (traced by galaxies participating in the mass
inflow) can give direct information about the mass profile (Kaiser 1987;
Reg\"os \& Geller 1989). 
Of course, $u(r)$ is not directly observable. Instead, for each galaxy we 
only observe its position on the sky, which translates into a projected 
distance from the assumed center of the 
structure, $r_\perp\leq r$, and its redshift, which can be translated into a
line-of-sight velocity $v$ with respect to the same center. 
The ``fundamental'' variables $r$ and $u$ and the ``observed'' variables
$r_\perp$ and $v$ (with the line-of-sight velocity of the structure's center
already subtracted) are related by

\begin{equation}
v=\pm\left[1-\left({r_\perp\over r}\right)^2\right]^{1\over 2}u(r).
\end{equation}

\noindent Contrary to the Hubble flow observed on large scales, 
$v$ is negative (approaching) for the more distant galaxies on the back side 
of each shell, and positive (receding) for the closer galaxies on the front
side. At any given projected distance $r_\perp$, one observes galaxies at
many different true distances $r$ from the center. The infall velocity
$u(r)$ decreases at large enough distances $r$, reaching zero at a finite 
(``turnaround'') radius $r_t$, and matching onto the Hubble flow, $u(r)=-Hr,$
for $r\gg r_t$. The projection factor increases from zero at $r=r_\perp$
(galaxies moving perpendicularly to the line of sight), asymptotically 
approaching unity for $r\gg r_\perp$. Therefore, there will be some maximum
projected velocity 

\begin{equation}
{\cal A}(r_\perp)\equiv\max_{r\geq r_\perp}|v(r_\perp,r)|=\max_{r\geq r_\perp}
\left[1-\left({r_\perp\over r}\right)^2\right]^{1\over 2}u(r),
\end{equation}

\noindent which is a monotonically decreasing function of $r_\perp$
(see Appendix), giving rise to caustics in the $(r_\perp,v)$ diagram 
with the characteristic ``trumpet shape'' 
described by Kaiser (1987) and by Reg\"os \& Geller (1989). 

In order to obtain the infall velocity $u(r)$ from the galaxy redshifts,
we first identify the caustics amplitude ${\cal A}(r_\perp)$ from the 
$(r_\perp,v)$ diagram by the procedure outlined in \S 3. Given this
relation, we can invert eq. (2) to obtain

\begin{equation}
u(r)\leq u_b(r)\equiv\min_{r_\perp< r}
{{\cal A}(r_\perp)\over\left[1-(r_\perp/r)^2\right]^{1\over 2}}.
\end{equation}

\noindent This is an {\it in}equality rather than an equality because for
an arbitrary shape of $u(r)$ it is {\it not} guaranteed that the shells
at {\it every} $r$ will correspond to a maximum amplitude for some $r_\perp$
(see Appendix for a detailed mathematical discussion).

If the mass density decreases outwards, then the collapse occurs from the 
inside out, with innermost mass shells first reaching 
turnaround and recontraction, and outer shells following in succession.
Any given shell will enclose the same mass $M$ at all times 
until it starts encountering
matter that has already passed through the center of the structure and is
again moving outwards. The latter is only expected to happen in the very
central, nearly virialized, part of the supercluster.
Elsewhere, the dynamics 
of any given shell is described by the well-known parameterized solution

\begin{equation}
r=A(1-\cos\eta);\quad t=B(\eta-\sin\eta); \quad A^3=GMB^2
\end{equation}

\noindent (e.g., Peebles 1993, chapter 20). 
Here, $A$ and $B$ are constants for any given shell (related to each other
by the enclosed mass $M$), and $\eta$ labels the ``phase'' of the shell's
evolution (initial ``explosion'' at $\eta=0$, maximum radius or ``turnaround''
at $\eta=\pi$, collapse at $\eta=2\pi$). As we are observing many shells
at one given cosmic time $t_1$ (measured from the Big Bang, at which all shells
started expanding, to the moment at which the structure emitted the light
currently being observed), for each shell we can write

\begin{equation}
{t_1\dot r\over r}=-H_0t_1{u\over H_0r}
={\sin\eta(\eta-\sin\eta)\over(1-\cos\eta)^2},
\end{equation}

\noindent where $H_0$ is the current value of the Hubble parameter.

We can determine an upper bound on $u(H_0r)$, and therefore on $u/(H_0r)$, by 
the procedure described above (note that $r$ itself depends on the
uncertainty in the cosmic distance scale). The combination $H_0t_1$ is
a dimensionless constant, dependent on the cosmological model (identified
by dimensionless constants such as the density parameters $\Omega$). 
At the redshift of the SSC ($z\approx 0.048$), it is likely to lie in the
range $0.62\leq H_0t_1\leq 0.95$, with the lower limit corresponding
to an Einstein-de Sitter Universe (with critical matter density, $\Omega_m=1$,
and no other ingredients), and the upper limit corresponding to an empty 
Universe ($\Omega_m=0=\Omega_\Lambda$) or a flat, low-density, 
$\Lambda$-dominated Universe ($\Omega_m=1-\Omega_\Lambda\approx 0.27$)
(e.g., Peebles 1993, chapter 13). For an assumed value of this parameter,
the left-hand side becomes fully determined, and the equation can be solved
for the value of $\eta$ for each shell. The equations can also be combined
to yield

\begin{equation}
H_0M(r)={(H_0r)^3\over G(H_0t_1)^2}{(\eta-\sin\eta)^2\over(1-\cos\eta)^3},
\end{equation}

\noindent the mass enclosed within the shell (of current radius $r$). 

For two reasons, {\it a mass estimate obtained by applying this model
to real data should be regarded as an upper bound on the true mass}. 
First, the model gives $u_b(r)$, an upper bound on $u(r)$, and this
upper bound is used to derive the mass. Second, the caustics amplitude 
${\cal A}(r_\perp)$ is amplified through random motions due to substructure 
within the infalling matter (Diaferio \& Geller 1997). Keeping this in mind,
we will use the model to put a bound on the mass of the collapsing region around 
A 3558.

\subsection{Diaferio's prescription}

On the other hand, Diaferio \& Geller (1997; see also Diaferio 1999)
have shown that the mass profile of structures forming in numerical 
simulations can be recovered to good precision from the formula

\begin{equation}
M(r)={{\cal F}_\beta\over G}\int_0^r{\cal A}^2(r_\perp)dr_\perp.
\end{equation}

There is no rigorous derivation for this result, although it can be
justified heuristically by assuming that ${\cal A}$ reflects the escape
velocity at different radii, i.e., that all galaxies within the caustics
are gravitationally bound to the structure. One has to assume further that the 
radial density profile lies between $\rho\propto r^{-3}$ and $\rho\propto r^{-2}$
(Diaferio \& Geller 1997; Diaferio 1999),
as in the outskirts of simulated clusters of galaxies (e.g., Navarro, Frenk, \&
White 1997). This mass estimate is independent of
the parameter $H_0t_1$, since no dynamical evolution is involved.
It has already been applied to the Coma cluster (Geller, Diaferio, \& Kurtz 
1999).

\section{Identifying the structure}

The $(r_\perp,v)$ diagram for the galaxies with available redshifts in the
Shapley region (see Paper II), shown in Fig. 1 for an adopted center at the 
main cluster A 3558 ($\alpha_c=13^h27^m56.9^s, \delta_c=-31^\circ 29'44'', 
cz_c=14300$ km s$^{-1}$), indeed shows the predicted ``trumpet shape'', 
with a maximum half-width ${\cal A}(r_\perp\approx 0)\sim 2000$ km s$^{-1}$,
and extending cleanly out to $r_\perp\sim 8 h^{-1}$ Mpc, and
less cleanly to perhaps $14 h^{-1}$ Mpc from the center.\footnote{At the 
redshift of A 3558, $z_c=0.048$, assumed to be of purely cosmological origin,
relative line-of-sight velocities are given by 
$v=c(z-z_c)/(1+z_c)=0.955c(z-z_c)$
(Harrison \& Noonan 1979). In an Einstein-de Sitter Universe ($q_0=1/2$), 
physical distances projected on the sky are given by 
$r_\perp=2c\theta[1-(1+z_c)^{-1/2}]/[H_0(1+z_c)]$
(e.g., Peebles 1993), where $\theta$ is the angular distance in radians.
Again at the redshift of A 3558, $1^\circ$ corresponds to 
$2.3 h^{-1}$ Mpc. The latter result is fairly insensitive to the choice
of cosmological model, therefore we adopt it in general.} Thus,
there is indeed a coherent structure
({\it ``extended core''}), enclosing at least (i.e., within $8 h^{-1}$ Mpc)
11 Abell clusters (Abell, Corwin, \& Olowin 1989) and three rich groups not 
included in the Abell catalog (see Table 1). 
All of these have relative velocities with
respect to A 3558 smaller than 1200 km s$^{-1}$, with a median 
$|v|_{med}\approx 200$ km s$^{-1}$.   
Given its shape, this extremely cluster-rich structure is most likely 
due to gravitational collapse.

Of course, the morphology of the SSC or even of its 
``extended core'' does not support the assumption of
spherical symmetry. The inner core (formed by A 3558, A 3556, A 3562, 
SC 1327-312, 
and SC 1329-314) is clearly elongated, and there is evidence that both the
galaxies and the clusters in the region of interest tend to form a planar
structure (Paper I; Bardelli et al. 2000; Paper II). This is supported by
the elementary fact that (within the extended core and a fair distance beyond 
it) all the clusters in the north (A 3559, A 3557a, A 3555,
further away A 1736b) have lower redshifts than A 3558, while those in the south 
(A 3560, A 3554, others at larger distances) and west (particulary those 
belonging to the {\it NW filament,} which extends towards A 3528 and is described
in Paper II) have higher redshifts. It is also hinted at by the slight upturn of the structure in 
Fig. 1 at $\theta^>_\sim 3.5^\circ$ (${r_\perp}^>_\sim 8 h^{-1}$ Mpc), which is 
due to the concentrations of clusters to the south-west and south-east of the 
central concentration. (The downward-pointing arm at $\theta\sim 4^\circ$
is due to the clusters A 1736a,b and A 3571.) 

However, 1) the extreme simplicity of this symmetry assumption 
compared to any possible improvement to it, 2) our ignorance of the
true 3-d distribution of {\it matter} in the supercluster, and 3) the fact
that the gravitational potential is generally more spherical than the
mass distribution originating it, motivate us to still apply a spherical model.
Note, in fact, that the model requires the gravitational potential
to be spherically symmetric, but no symmetry assumptions are being made about the
distribution of galaxies as tracer particles. 

Although the structure appears well-defined to the eye, it does not have a
perfectly sharp edge, as it would be expected to have in the perfect 
spherical infall model. Departures from spherical symmetry, substructure
such as clusters, the finite number of ``test particles,'' and 
(to a much lesser degree) observational uncertainties 
wash out and distort the structure. This makes a determination of the caustics
non-trivial. In the rest of this section, we follow Diaferio's (1999) 
general approach in first smoothing the data, i.e., obtaining a smooth
estimate $f(r_\perp,v)$ for the density of observed galaxies on the $(r_\perp,v)$
plane, and then applying a cut at some density contour which is taken to 
correspond to the caustics.
The details of how each of these steps is carried out differ slightly from
Diaferio's approach, and are discussed in the rest of this section.

\subsection{Density estimation in the $(r_\perp,v)$ diagram}

For a global analysis of the central (collapsing) region of the
SSC, we need to obtain a smooth estimate $f(r_\perp,v)$ of the
density of galaxies in the $(r_\perp,v)$ diagram. Density estimation has been
discussed by many authors, such as Silverman (1986), and in the astronomical
context Pisani (1993; 1996) and Merritt \& Tremblay (1994). 

Diaferio (1999) applied density estimation to the particular 
problem of interest here. For $N$ data points (galaxies) with coordinates
$(r_\perp^i,v^i)$, he adopts the estimate

\begin{equation}
f(r_\perp,v)={1\over N}\sum_{i=1}^N{1\over h_r^i h_v^i}
K\left({r_\perp-r_\perp^i\over h_r^i},{v-v^i\over h_v^i}\right),
\end{equation}

\noindent where

$$K(\vec t)=\cases{4\pi^{-1}(1-|\vec t|^2)^3 &if $|\vec t|<1$,\cr
0, &otherwise,}$$

\noindent is a smooth, but centrally peaked, kernel function. 
The ratio of smoothing lengths, $q=h_v^i/h_r^i$ is fixed, approximately equal
to the ratio of observational uncertainties 
($q=50 {\rm km\ s}^{-1}/0.02 h^{-1} {\rm Mpc}=25 H_0$), and the individual values
of, say, $h_r^i,$ are chosen by an adaptive algorithm.

Any choice of smoothing lengths is a compromise between keeping as much 
structure as possible (favoring small smoothing lengths) while eliminating
as much noise as possible (favoring large values). The ideal compromise,
though quite subjective in any case, depends on the density of data points,
which generally varies over the volume being studied, motivating the 
choice of a different smoothing length for each data point. For our particular 
application, the density of points does not vary enormously over the area
of interest, and we are only interested in the overall envelope of the
structure, not in fine details. Therefore, we consider the additional 
computational effort of adaptive smoothing with different local smoothing
lengths unjustified. We apply fixed, overall smoothing lengths 
$h_r=1 h^{-1}$ Mpc, $h_v=$ 500 km s$^{-1}$ (giving $q=5H_0$),
chosen by eye to preserve the
overall shape while minimizing the noise, and each corresponding to about
1/8 of the total extension of the structure studied. Changing either of the 
two lengths by a factor of 2 either way does not substantially change our 
results. We note also that the chosen lengths are much larger than the 
respective uncertainties in the data, which therefore become irrelevant in
determining the detected structures. 

One problem with the smoothing kernel given above is that the data have a natural
cutoff at $r_\perp=0$, where they go abruptly from a fairly high (near maximum) 
density (at $r_\perp>0$) to zero (at $r_\perp<0$). When applied to points of 
small (positive) $r_\perp$, the smoothing kernel
extends to negative values (where there are no data points), producing a 
decrease in $f(r_\perp,v)$ when approaching $r_\perp=0$ from above. This causes
isodensity contours to narrow as $r_\perp\to 0^+$, as seen, e.g., in Fig. 1 
of Geller, Diaferio, \& Kurtz (1999), in Figs. 4 and 5 of Diaferio (1999), and 
in Fig. 2a of the present paper. 

This can be cured, e.g., by making a mirror image of the data at $r_\perp<0$ and 
letting the smoothing kernel integrate over both the real data and their image 
(Fig. 2b). Aside from correcting for the ``misbehavior'' at $r_\perp=0$, this 
procedure gives results very similar to that of Diaferio (1999).

A more rigorous approach is suggested
by Merritt \& Tremblay (1994), who deal with density estimation in circularly 
symmetric structures. They focus on the surface density $\Sigma(r_\perp)$
(number per unit area) rather than the radial density $N(r_\perp)=2\pi r_\perp\Sigma(r_\perp)$ 
(number per unit radial coordinate), and estimate $\Sigma(r_\perp)$ with a 
circularly averaged kernel. In our case (with one additional coordinate $v$,
which is not affected by this problem), we can define a number density per
unit (projected) area per unit line-of-sight velocity as 
$\Sigma(r_\perp,v)=f(r_\perp,v)/(2\pi r_\perp),$ and estimate it through
a kernel which is a product of a standard, one-dimensional kernel for
$v$ and a circularly averaged kernel for $r_\perp$. In particularly, Fig. 2c
shows the results of applying to our data a one-dimensional quadratic 
({\it Epanechnikov}) kernel for $v$, and an annularly averaged, two-dimensional 
quadratic kernel for $r_\perp$. (See Merritt \& Tremblay 1994, eqs. 9a and 28a
for explicit formulae.) The density $\Sigma$
obtained from this procedure is well-behaved in all respects, decreasing
from $r_\perp=0$ outwards. Indeed, it decreases so quickly that the density
contours tend to close at fairly small radii, contrary to the visual impression
from the data. Therefore, these contours are unlikely to be realistic representations
of the velocity caustics. 

A final alternative (with the added virtue of reducing biases due to non-uniform 
spatial sampling) is to normalize $f(r_\perp,v)$ at each given $r_\perp$ with 
respect to the value at $v=0$, i.e., take a density estimate 

\begin{equation}
\tilde f(r_\perp,v)\equiv{f(r_\perp,v)\over f(r_\perp,0)},
\end{equation} 

\noindent with the ``original'' $f(r_\perp,v)$ determined by any of the other
methods (the case shown in Fig. 2d is based on Diaferio's estimator). This 
estimator gives results very similar to the first two. 

Overall, we consider that the second procedure (the ``mirror image'' density
estimate)
is the one that most closely represents the visual appearance of the data,
while at the same time having mathematically desirable properties (smooth
density contours slowly narrowing with increasing $r_\perp$), and being close
enough to Diaferio's to permit a direct comparison of results. Therefore, we 
use the ``mirror image'' density estimation for the analysis that follows. 
However, we stress that it is a very arbitrary choice, and that other choices 
may give quite different final results. However, discarding the very different
result based on the ``surface density'' scheme, the other
procedures discussed above give masses that, at any given radius, differ by 
less than 10 \% from the one obtained by the ``mirror image'' procedure. 

\subsection{Finding the caustics}

Given the estimated density $f(r_\perp, v)$, we now turn to finding the 
caustics which separate the collapsing structure from the galaxies in the
foreground and background. We are again inspired by Diaferio (1999), who
uses a fixed density cutoff, $f(r_\perp, v)=\kappa$, with the value of
$\kappa$ fixed by virial arguments applied to the most central region.
In principle, taking a fixed value is somewhat arbitrary. By plotting in 
$(r_\perp,v)$ space, we are summing galaxies over annuli, and therefore
a uniform background of galaxies would result in a density increasing 
$\propto r_\perp$, and therefore a fixed cutoff might include more and
more of the background as $r_\perp$ increases. The problem is worsened by
the non-uniformly sampled data in our particular case. Therefore, a
more natural and in principle better way to distinguish the structure from
the background might be to fix on maxima of ${\partial f\over\partial v}$
for each $r_\perp$. However, taking and maximizing a {\it derivative} of 
the numerically determined function is much noisier than just imposing 
a fixed cutoff. Therefore, we follow Diaferio in adopting the latter approach.

In order to choose the value of the cutoff, Diaferio (1999) uses the
virial theorem to relate the escape velocity (in his interpretation 
represented by the velocity amplitude ${\cal A}$) and the velocity dispersion
$\sigma$ of the central cluster, therefore writing 
$\langle{\cal A}^2\rangle_{\kappa,R}=4\sigma^2$, where the average is a galaxy
number-weighted average over the region enclosed by the virial radius $R$
of the central cluster, for a given density cutoff $\kappa$. For A 3558,
the velocity dispersion is a relatively robust number 
($\sigma\approx 928$ km s$^{-1}$), not sensitive to the
radius of the sphere to be averaged over, and in the central region ${\cal A}$
(determined with any reasonable cutoff) is also fairly radius-independent.
Therefore, the dependence on the (poorly determined) virial radius is weak, 
and we arbitrarily choose it as $R=1 h^{-1}$ Mpc, and do a straight radial
average ({\it not} number-weighted) to calculate 
$\langle{\cal A}^2\rangle_{\kappa,R}$ and determine $\kappa$ by Diaferio's
condition. 
 
We tested the validity of Diaferio's condition by the following procedure.
Figure 3 shows the area of the $(r_\perp,v)$ diagram enclosed by contour 
levels with different $\kappa$. We clearly distinguish 3 regimes: 

\begin{itemize}
\item[(a)] At very low densities, the enclosed area is most of the diagram,
therefore enclosing much of the background, not belonging to the structure.
The area rapidly decreases as the threshold density is increased.

\item[(b)] At intermediate densities, the decreasing curve becomes much 
flatter ($dA/d\kappa\approx$ constant), and we interpret this as having most 
of the background excluded and probing progressively denser parts of 
the structure. This is confirmed by watching the contour plots, which indeed
trace the boundaries of the structure, and become progressively tighter.

\item[(c)] Finally, at high values of $\kappa$, only a few isolated peaks in
the structure are left enclosed, and these finally disappear when $\kappa$ 
reaches the maximum density present.
\end{itemize}

This analysis suggests choosing the threshold at the transition between regimes
(a) and (b). This falls close to the threshold value chosen by Diaferio's
(1999) condition as outlined above, marked by the vertical line.
This strengthens the argument for Diaferio's choice of cutoff, 
already used to choose the contours in Fig. 2, and adopted hereafter. 

The chosen density contour of course gives two values of $v$ (one positive, 
$v_u$, and one negative, $v_d$) for each value of $r_\perp$. In the SSC,
it turns out that the upper contour is much ``cleaner'' (separating a dense
region from a nearly empty one), therefore we simply adopt ${\cal
A}(r_\perp)=v_u(r_\perp)$ rather than Diaferio's prescription 
${\cal A}(r_\perp)=\min\{|v_u(r_\perp)|,|v_d(r_\perp)|\}.$

\section{Results and Discussion}

Fig. 4 shows the enclosed mass as a function of radius, $M(r)$, as determined
by the two methods discussed in \S 2, together with a third determination,
namely the cumulative mass of the clusters enclosed in the given radius,
given in Table 1. Note that the mass estimates $M_{500}$, taken from Ettori et 
al. (1997) for the most important clusters, are masses within a radius 
enclosing an average density 500 times the critical density $\rho_c$. This is 
substantially higher than the standard ``virialization density'' of
$\sim 200\rho_c$, and therefore gives a 
conservative lower limit to the total virialized mass, which may be
increased by a factor $\sim (500/200)^{1/2}\approx 1.58$ for a more realistic
estimate. 

Several comments are in order: 

1) As discussed above, the pure spherical infall model is highly idealized
and, even if correct, can only give an upper bound on the mass within any given
radius. Therefore, the upper (dot-dashed) curve, corresponding to pure
spherical collapse in a cosmological model with $H_0t_1=0.62$ 
should be regarded as a fairly robust upper limit to the mass within any given
radius.

2) The model of Diaferio \& Geller (1997) has been calibrated against 
simulations. Applied to the infall regions around clusters of galaxies,
it should in principle give the correct mass to within about
$25 \%$ (Geller et al. 1999). However, it assumes a density profile 
decreasing at least as fast as $\rho(r)\propto r^{-2}$, which
may not apply to the very noisy region around A 3558, which contains a 
number of other, fairly rich clusters. It appears surprising that the mass
profile it gives for this region is quite similar (both in shape and in 
amplitude) to that obtained by Geller et al. (1997), considering that 
Coma is a quite massive cluster (as massive or perhaps even more massive 
than A 3558), but is {\it not} surrounded by any other massive clusters. 

3) The mass estimate based on individual cluster masses is uncertain
for two reasons. First, of course it does not consider the mass in the
non-virialized outskirts of the clusters or not associated with clusters
at all, and therefore it would be expected to underestimate the total
mass. On the other hand, in the absence of information on the three-dimensional
distance $r$ of each cluster to A 3558, and given that the velocity does not
give reliable distance information within the collapsing structure, 
each cluster was put at its projected radius $r_\perp\leq r$, and therefore
contributes to the enclosed mass already at radii smaller than its true
position. Therefore, the mass in virialized clusters within any given
radius $M(r)$ is overestimated by the projection into radius $r$ of clusters
actually at larger radii. 

Given these caveats, there seems to be fair agreement among the different mass
determinations, and it seems safe to say that the mass enclosed by radius
$r=8 h^{-1}$ Mpc lies between $2\times 10^{15}$ and 
$1.3\times 10^{16} h^{-1}M_\odot$. It is interesting, nevertheless, that
Diaferio's method gives results that differ so little from the lower limit
to the virialized mass in clusters. Therefore, if Diaferio's method is
applicable to the SSC, the either there would be very little mass outside 
the inner, virialized parts of clusters of galaxies in this region, or the 
cluster mass estimates would have to be systematically high.

For comparison, Ettori et al. (1997) used three different mass estimates,
namely: 1) the sum of the gravitational masses of clusters as obtained from
their X-ray emission profile, $M_{grav}$, 2) the total mass expected to be associated
with the baryons observed in clusters, $M_{PN}$, and 3) the mass obtained from
applying the virial theorem to the enclosed clusters, used as test particles,
$M_{vir}$. They applied these methods to four progressively larger structures, 
each enclosing the previous one. The one most similar to our $8 h^{-1}$ Mpc
sphere appears to be the second, enclosing 12 clusters, and with a nominal 
3-d radius of $13.9 h^{-1}$ Mpc, obtained by treating redshift as a third
coordinate, which we have argued to give an overestimated depth in the 
collapsing region. For this region, they find $M_{grav}=2.15$,
$M_{PN}=5.2\Omega_m$ and $M_{vir}=1.75$, all in units of $10^{15}h^{-1}M_\odot$.
The first two are likely underestimates (as they consider only the matter in
the virialized regions of clusters observed in X-rays), and the third
is completely uncertain, given the uncertain distances along the line of sight
and the fact that the SSC is not virialized (but see Small et al.
1998 for a modern, more careful application of the virial theorem to the
Corona Borealis supercluster). Thus, it is reasonable that we find a somewhat
higher mass for the (optically observed) clusters, and a possibly much
higher total dynamical mass, as suggested by the spherical collapse model.
 
The average enclosed density (see Fig. 5) drops from a value 400 --
500 times the critical density within $1 h^{-1}$ Mpc (consistent with the
presence of a massive, already collapsed and virialized cluster) to 
a value still several times critical ($\sim 3.7$ to 23 times, depending on
the model) within our outermost radius, $8 h^{-1}$ Mpc. From galaxy counts
in redshift space, Bardelli et al. (2000) find an overdensity 
$N/\bar N=11.3\pm 0.4$ within a region of equivalent radius $10.1 h^{-1}$ Mpc. 
Assuming that galaxies trace mass, this might in principle allow us to 
determine the universal matter density parameter  
$\Omega_m=\bar\rho/\rho_{\rm crit}=(\rho(r)/\rho_{\rm crit})/(N(r)/\bar N).$
In practice, however, the uncertainty in this estimate is still much 
too large to put a useful constraint on $\Omega_m$.

Fig. 6 shows that, in the spherical collapse model, the whole structure within
$8 h^{-1}$ Mpc has already been contracting for more than 1/3 of its lifetime, 
and the inner regions are in the final stages of collapse,
consistent with the presence of a massive cluster. Even if there were no
additional mass beyond $8 h^{-1}$ Mpc, the current turnaround radius would be
at $\sim 14 h^{-1}$ Mpc, and the bound region (to collapse eventually) would
extend to $\sim 20 h^{-1}$ Mpc, enclosing essentially the whole supercluster, 
including the strong concentration around A 3528, A 3530, and A 3532.
However, the much lower enclosed densities in the Diaferio \& Geller model 
would imply that the $8 h^{-1}$ region around A 3558 is (at best) only now 
reaching turnaround, and has another Hubble time to go for final collapse. 

As discussed in Paper I, the mass required at the distance of the
SSC to produce the observed motion of the Local Group with
respect to the cosmic microwave background is given by

$$M_{dipole}\approx 4.5\times 10^{17}\Omega_m^{0.4}h^{-1}M_\odot.$$

\noindent The mass within $8h^{-1}$ Mpc can therefore produce at most 
$3\Omega_m^{-0.4}\%$ of the observed Local Group motion, which makes
it unlikely that even the whole SSC would dominate its
gravitational acceleration. 

Unfortunately, not much can be said about the regions beyond a radius 
of $\sim 8 h^{-1}$ Mpc from the center. It seems safe to assert, though,
that the collapsing region does not extend far beyond, say, $14 h^{-1}$ Mpc.
This gives an upper bound on the average density enclosed in larger radii,
$\bar\rho<3\pi/(32Gt_1^2)$. Thus, in order to produce the peculiar 
velocity of the Local Group, one would need a region of radius

$$r>55h^{-1}\Omega_m^{0.13}(H_0t_1)^{2/3}{\rm Mpc},$$

\noindent which, for the range of cosmological values considered before,
corresponds to a lower limit $\sim 40 h^{-1}$ Mpc. Therefore, in the
unlikely case that the whole SSC (characterized in Paper II)
were on the verge of gravitational collapse, it would be able to produce 
on its own the observed peculiar velocity of the Local Group. This statement
ignores, of course, that the apex of the Local Motion does not point exactly
at the SSC, and therefore some additional contribution is necessary in any
case.

\section{Conclusions}

We have presented the (to our knowledge) first application of a plausible
dynamical model to a supercluster of galaxies, containing a substantial
number of clusters. The central $8 h^{-1}$ Mpc region 
of the Shapley spercluster (and probably a much more extended region 
surrounding it) is argued to be currently collapsing under the
effect of its own gravity. Its mass, although uncertain due to 
idealizations in the model, indicates a large enhancement over  
the average density of the Universe, although still far from that required
to produce the Local Group's observed motion with respect to the cosmic
microwave background.

\acknowledgements

The authors thank A. Diaferio for interesting discussions at the First 
Princeton-U. Cat\'olica Astrophysics Workshop, {\it The Cosmological Parameters
$\Omega$,} held in Puc\'on, Chile, in January 1999, and for extensive e-mail
exchanges thereafter. A previous version of the kernel smoothing program
used here was due to A. Meza. We also thank him and R. Benguria for useful 
discussions.
This work was financially supported by FONDECYT grant 8970009 ({\it Proyecto
de L\'\i neas Complementarias}), and by a Presidential Chair in 
Science awarded to H. Quintana. E.R.C. was funded by FAPESP Ph.D. fellowship 
96/04246-7.

\appendix
\section{Geometric interpretation and mathematical properties of the
line-of-sight velocity amplitude ${\cal A}(r_\perp)$ and the infall 
velocity $u(r)$}

In order to derive and understand intuitively the properties of the
observed velocity amplitude ${\cal A}(r_\perp)$ and the infall velocity
$u(r)$ producing it in the spherical model, it is convenient to
define new variables $P=r_\perp^2$, $R=r^2$, $V={\cal A}^2$, and $U=u^2$.
Then, 

\begin{equation}
V(P)=\max_{R>P}F(P,R), \qquad {\rm with}\quad 
F(P,R)\equiv\left(1-{P\over R}\right)U(R).
\end{equation}

\noindent (We consider only the interval in which $u(r)=-\dot r\geq 0,$
corresponding to infall.)

Note that, seen as a function of $P$ for given $R$, $F(P,R)$ is a
straight line intersecting the horizontal axis at $P=R$ and the vertical
axis at $F(0,R)=U(R)$. Therefore, the function $U(R)$ defines a set of
straight lines whose upper envelope gives the function $V(P)$ (see Fig. 7). 
The relation between the functions $U(R)$ and $V(P)$ is very similar to
the {\it Legendre transformation} (e.g., Courant \& Hilbert 1989, \S I.6), 
and shares many of its properties. 

Since $U(R)\geq 0$ for all $R$, {\bf $V(P)$ is positive ($V\geq 0$), monotonically
decreasing ($V'\leq 0$), and convex ($V''\geq 0$).} Fig. 8 shows the function
$V(P)$ obtained from the data in the way described in this paper. It is clear
that it does not strictly satisfy the conditions of monotonicity and convexity,
indicating that, as expected, the pure spherical infall model does not 
exactly represent the data.

We can establish a relation $R(P)$ in the sense
that, for any given $P=P_0$, $R_0\equiv R(P_0)$ is (are) the value(s) of 
$R$ for which the maximum of $F(P_0,R)$ occurs, so that 

\begin{equation}
V(P_0)=F(P_0,R_0)=\left(1-{P_0\over R_0}\right)U(R_0)
\end{equation}

\noindent (see Fig. 7). In addition, the linear function $F(P,R_0)$ is the tangent
to $V(P)$ at $P_0$, so

\begin{equation}
V'(P_0)={\partial F\over\partial P}(P_0,R_0)=-{U(R_0)\over R_0}.
\end{equation}

$R(P)$ is a strictly increasing function, but it is not necessarily 
continuous. For example, it can happen that, for some point $P_0$, the 
maximum occurs at the intersection of two straight lines labeled by $R=R_1$ 
and $R=R_2>R_1$ (see Fig. 9), with the lines corresponding to all
other values of $R$ lying below it, i.e., 

\begin{equation}
V(P_0)=\left(1-{P_0\over R_1}\right)U(R_1)=\left(1-{P_0\over R_2}\right)U(R_2)
\geq \left(1-{P_0\over R}\right)U(R) \qquad \forall R \in (R_1,R_2).
\end{equation}

\noindent The values of $U(R)$ in the open interval $(R_1,R_2)$ do not affect
the function $V(P)$ and therefore cannot be recovered from it. In general,
it can only be said that

\begin{equation}
U(R)\leq U_b(R)\equiv\min_{P<R}{V(P)\over 1-P/R}.
\end{equation}

\noindent This is an equality for those $R$ which correspond to a maximum 
$F(P,R)$ for some $P$, and a strict inequality in all other cases. The latter
can in principle be diagnosed by realizing that, in the case discussed above, 

\begin{equation}
\lim_{P\to P_0^-}V'(P)=-{U(R_1)\over R_1}
<\lim_{P\to P_0^+}V'(P)=-{U(R_2)\over R_2},
\end{equation}

\noindent so both $R(P)$ and $V'(P)$ are discontinuous at $P=P_0$. In practice,
with noisy data, a discontinuity in $V'(P)$ is difficult to detect, and the
inequality, eq. (A5), has to be used as such. It is interesting to note that,
taking the reciprocal value of all variables, one can write

\begin{equation}
U^{-1}(R^{-1})\geq U_b^{-1}(R^{-1})
=\max_{P^{-1}>R^{-1}}\left(1-{R^{-1}\over P^{-1}}\right)V^{-1}(P^{-1}).
\end{equation}

\noindent This has the same form as eq. (A1). We can conclude that 
$U_b^{-1}(R^{-1})$ has the same properties as $V(P)$, being positive,
monotonically decreasing, and convex, in fact, it is the convex and decreasing 
lower envelope of $U^{-1}(R^{-1})$. As long as the latter is itself convex
and decreasing, then $U_b(R)=U(R)$, while in general $U_b(R)\geq U(R)$.
Fig. 10 shows $U_b^{-1}(R^{-1})$ as obtained from the data. Its curved parts 
(here absent) and ``corners'' are expected (within the pure spherical 
collapse model)
to correctly estimate $U^{-1}(R^{-1})$, while the straight segments are
lower bounds.

\newpage

\figcaption[fig1.ps]{Distribution of galaxies and clusters of galaxies in the Shapley 
Supercluster, with reference to the central cluster A 3558. The abscissa 
represents the angular distance $\theta$ of each object to the center of A 3558 
(at $\alpha_c=13^h27^m56.9^s, \delta_c=-31^\circ 29'44''$). The vertical axis is 
the line-of-sight velocity, $cz$, where $c$ is the speed of light and $z$ is 
the redshift. Dots are individual galaxies, circles represent the centers of
clusters and groups of galaxies (see Papers II and IV). Note the dense, 
``trumpet shaped'' 
region extending horizontally from the location of A 3558 (circle at $\theta=0$, 
$v\approx 14300$ km s$^{-1}$), which is interpreted as the collapsing 
structure.\label{fig1}}

\figcaption[fig2.ps]{The four panels show the central part of the diagram in 
Fig. 1, with line-of-sight velocities expressed with respect to A 3558, 
with the projected radius $r_\perp$ as the abscissa, and with an isodensity 
contour superimposed. Each panel corresponds to a different density
estimation scheme. In panel (a), the density estimate is a standard 
two-dimensional (fixed) kernel smoothing; in (b)
it is modified by considering a ``mirror image'' of the data to the left-hand 
side of the vertical axis; in (c) the estimate is the circularly averaged 
``surface density'' estimate based on Merritt \& Tremblay (1994); in (d)
the estimate is the same as in (c), normalized to the values on the 
horizontal axis. More details on each of these are given in \S 3.1, and the 
choice of isodensity contours is discussed in \S 3.2.\label{fig2}}

\figcaption[fig3.ps]{Area of the $(r_\perp,v)$ diagram enclosed by 
isodensity contours $f(r_\perp,v)=\kappa$, as a function of the density cut 
$\kappa$. The vertical line marks the density cut chosen as suggested by 
Diaferio (1999) and adopted in the present paper.\label{fig3}}

\figcaption[fig4.ps]{Enclosed mass as a function of radius around A 3558, 
given by different methods. The solid line is is the sum of 
masses of clusters and groups within the given radius, taking 
their projected
distance as the true distance to A 3558. The dashed and dot-dashed lines 
are upper bounds to the total mass based on the pure spherical infall
model, for $H_0t_1=0.95$ and $H_0t_1=0.62$, respectively. The solid-dotted
line is the estimate from Diaferio \& Geller's (1997) escape-velocity
model.\label{fig4}}

\figcaption[fig5.ps]{Logarithm of the average enclosed density in units of the 
critical density, as a function of radius. The symbols indicate the
models, as in Fig. 4.\label{fig5}}

\figcaption[fig6.ps]{Turnaround time in units of the 
current age of the Universe, as a function of radius. The symbols indicate the
models, as in Fig. 4.\label{fig6}}

\figcaption[fig7.ps]{Schematic illustration of the transformation relating
$V(P)={\cal A}(r_\perp^2)$ to $U(R)=u^2(r^2)$. See Appendix for an 
explanation.\label{fig7}}

\figcaption[fig8.ps]{The function 
$V(P)={\cal A}(r_\perp^2)$ determined from the data.\label{fig8}}

\figcaption[fig9.ps]{Schematic illustration of the transformation relating
$V(P)={\cal A}^2(r_\perp^2)$ to $U(R)=u^2(r^2)$, when there is a discontinuity
in the relation $R(P)$. See Appendix for an 
explanation.\label{fig9}}

\figcaption[fig10.ps]{The function $U^{-1}(R^{-1})=u^{-2}(r^{-2})$ 
determined from the data.\label{fig10}}

\newpage

\begin{deluxetable}{cccccc}
\tablecolumns{9}
\tablewidth{0pt}
\tablecaption{Clusters and groups of galaxies in the collapsing region around 
A 3558 ($r_\perp<8 h^{-1}{\rm Mpc}$, $|v|<2000{\rm km\ s}^{-1}$) }
\tablehead{
\colhead{Name}  & \colhead{$r_\perp$} & \colhead{$v$} & \colhead{$M_{500}$} &
\colhead{$\sum M_{500}$}          &  \colhead{Source}     \\
\colhead{} & \colhead{[$h^{-1}$ Mpc]} & \colhead{[km s$^{-1}$]} & 
\colhead{[$10^{14}h^{-1}M_\odot$]} &
\colhead{[$10^{14}h^{-1}M_\odot$]} & for $M_{500}$}\startdata
A 3558      & 0.00 &    0 & 6.09 &  6.09 & 1 \\     
SC 1327-312 & 0.94 & +700 & 2.14 &  8.23 & 1 \\
A 3556      & 1.92 &  -16 & 1.82 & 10.05 & 1 \\
SC 1329-314 & 1.93 & -956 & 1.03 & 11.08 & 1 \\
A 3562      & 2.94 &  -12 & 3.06 & 14.14 & 1 \\
A 3560      & 4.36 & +144 & 2.72 & 16.86 & 1 \\
A 3552      & 4.48 &+1153 & 0.36 & 17.22 & 2 \\
A 3559      & 4.65 & -376 & 0.83 & 18.05 & 1 \\
A 3554      & 6.13 &  +99 & 1.26 & 19.31 & 2 \\
A 3557a     & 6.18 & -114 & 0.47 & 19.78 & 2 \\
A 3555      & 7.07 & -419 & 0.10 & 19.88 & 2 \\
SC 1342-302 & 7.77 & +119 & 0.20 & 20.08 & 2 \\
A 724S      & 7.86 & +367 & 0.88 & 20.96 & 2 \\
A 726S      & 7.89 & +206 & 0.59 & 21.55 & 2 \\
\enddata
\tablecomments{Columns: 1. Cluster identification. 
2. Projected distance from A 3558. 
3. Average line-of-sight velocity, relative to A 3558. 
4. Adopted cluster mass within a radius enclosing an average density 500 times
the critical density, from X-ray or optical observations. 
5. Cumulative sum of cluster masses. 
6. Source of the cluster mass estimate: 
(1) X-ray deprojection analysis of Ettori et al. 1997;
(2) preliminary mass estimate from optical determination to be refined in
Paper IV.
(The entries in columns 2 and 3 were calculated from the data given in Paper 
II.)}
\end{deluxetable}

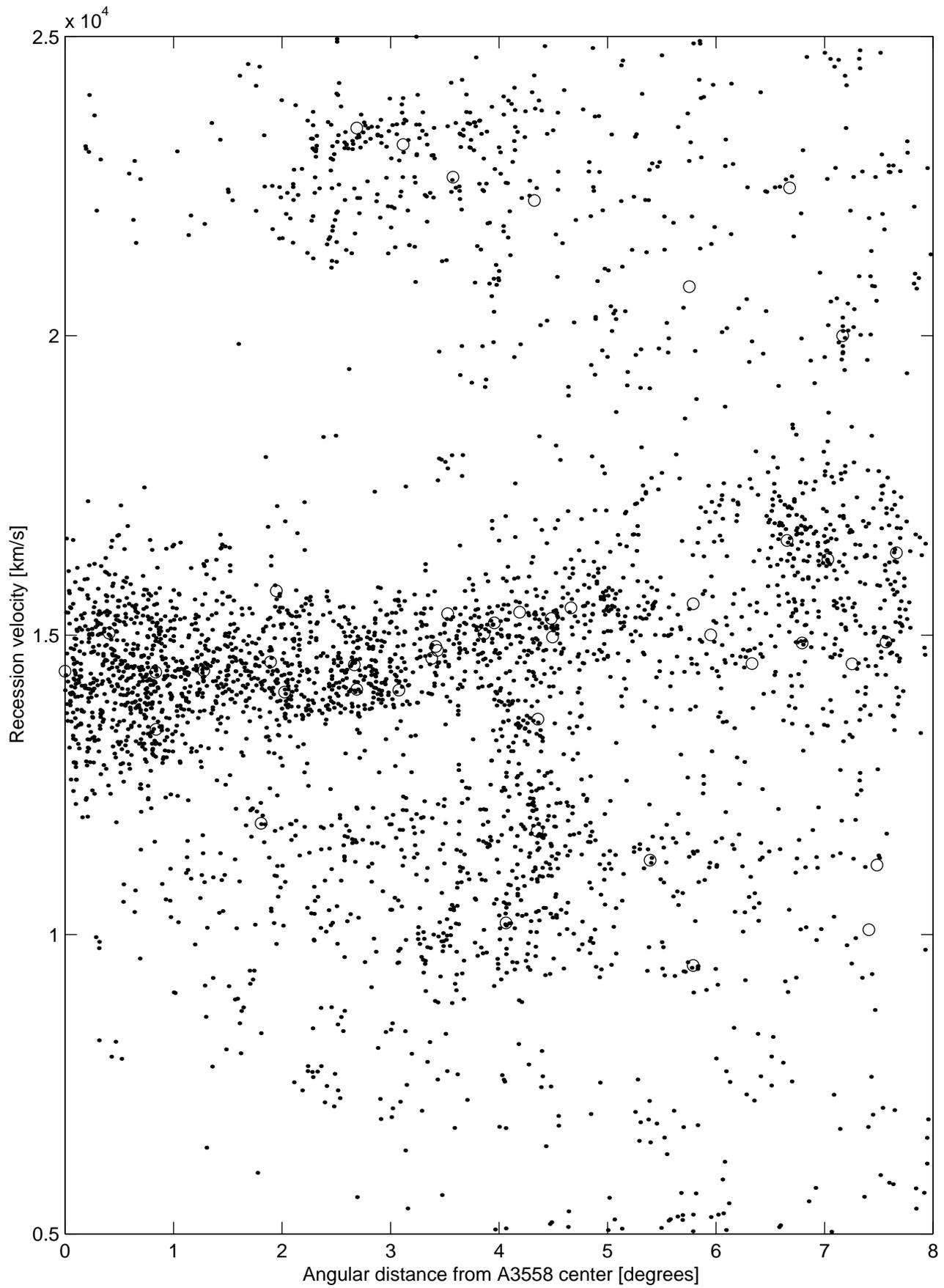
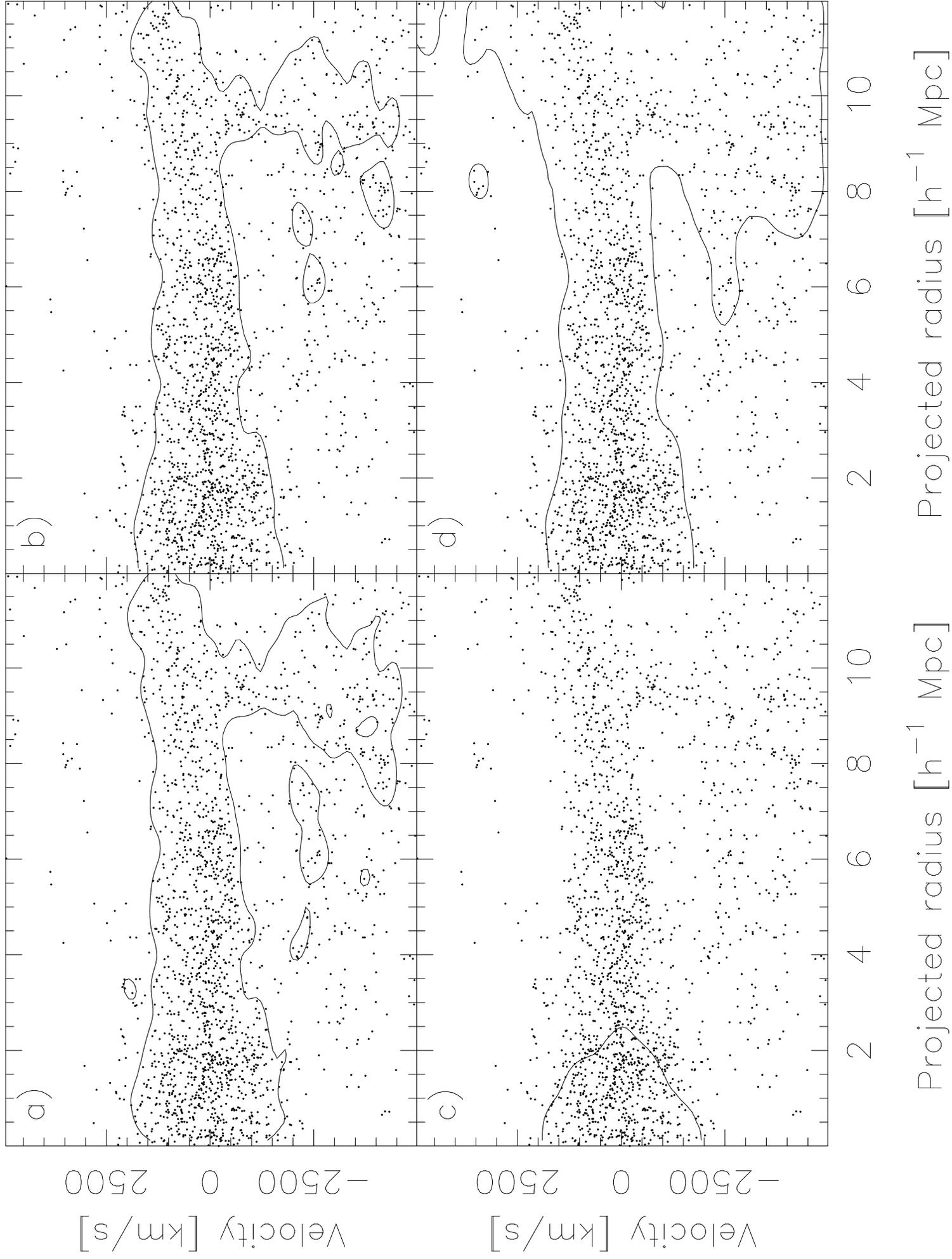
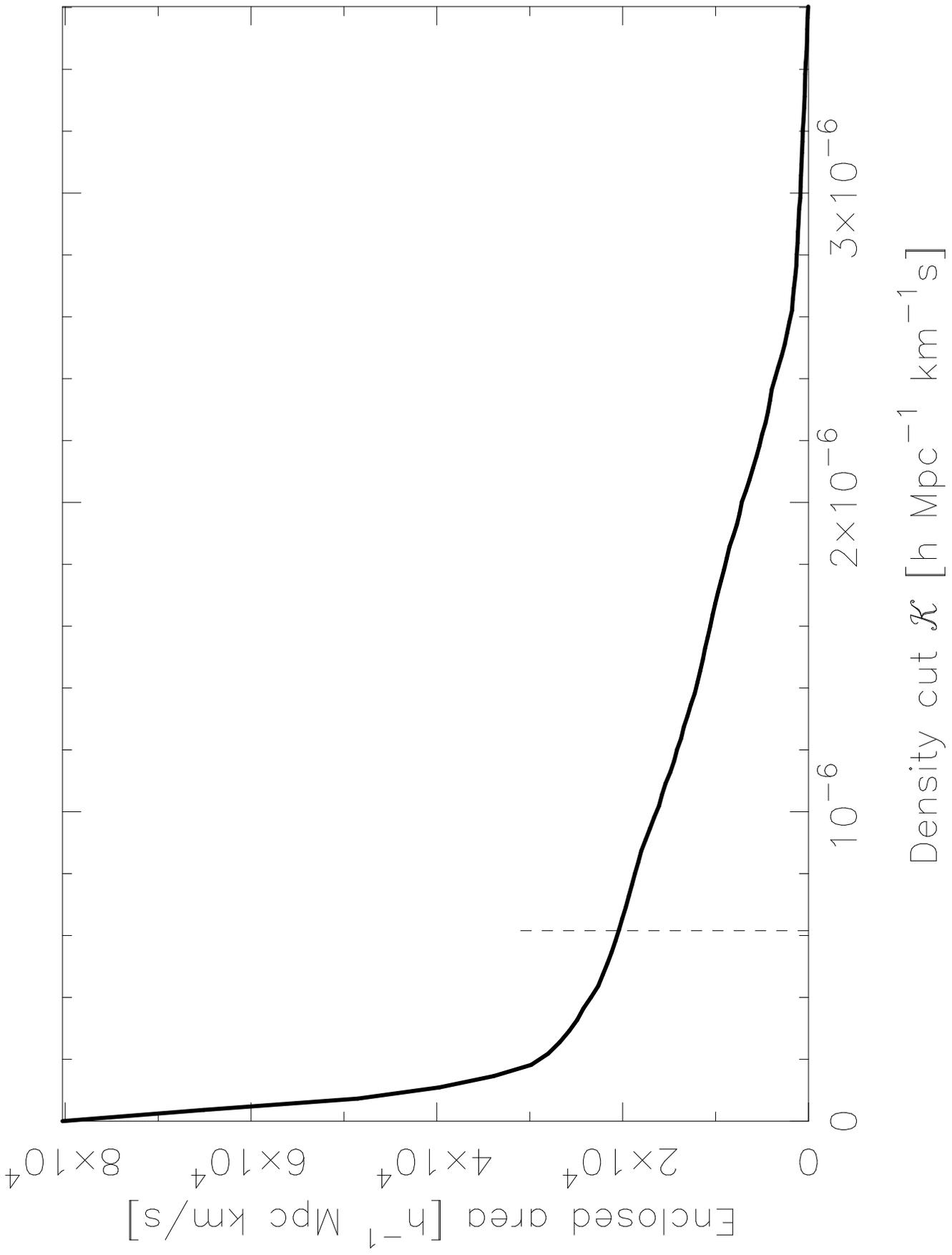
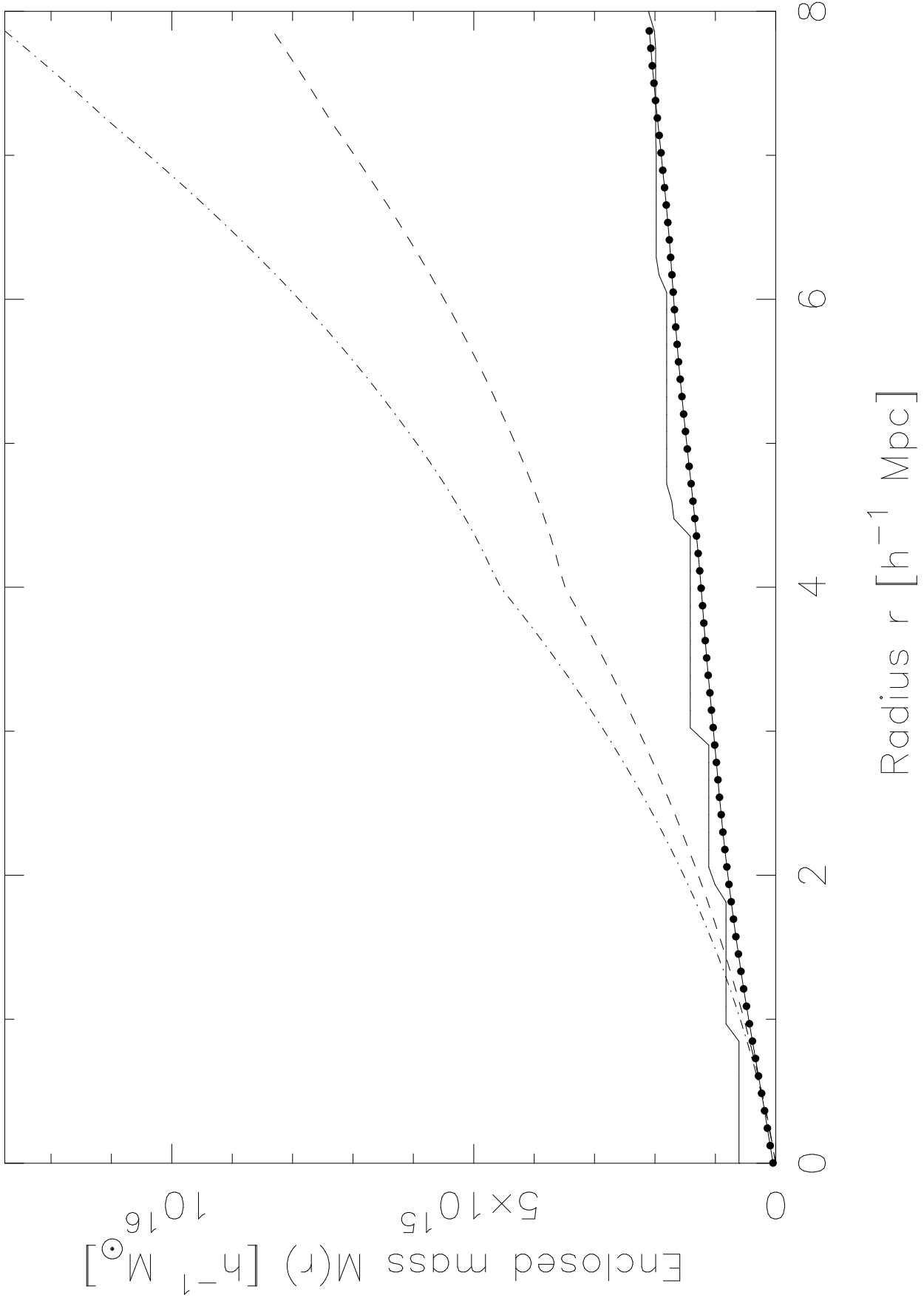
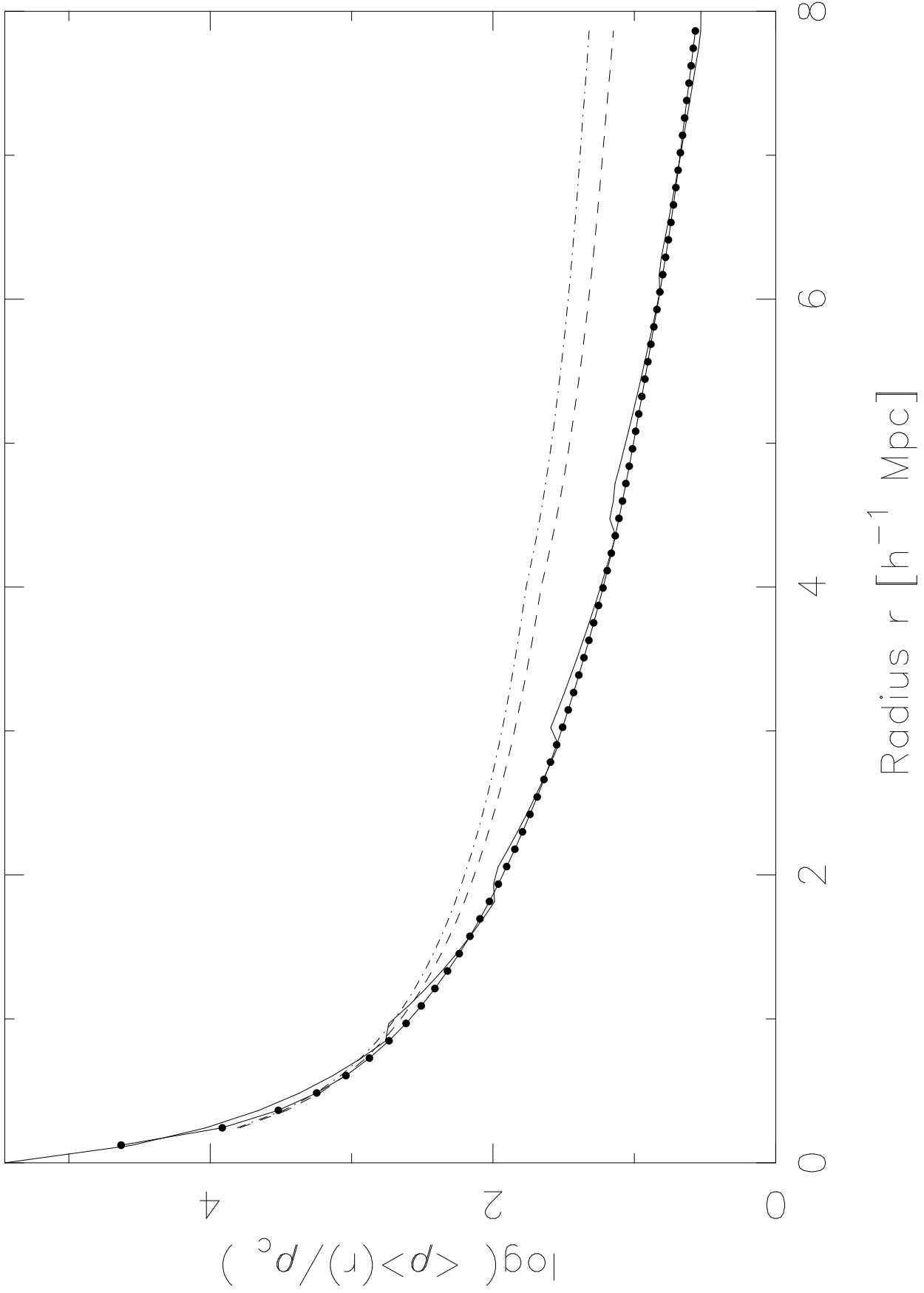
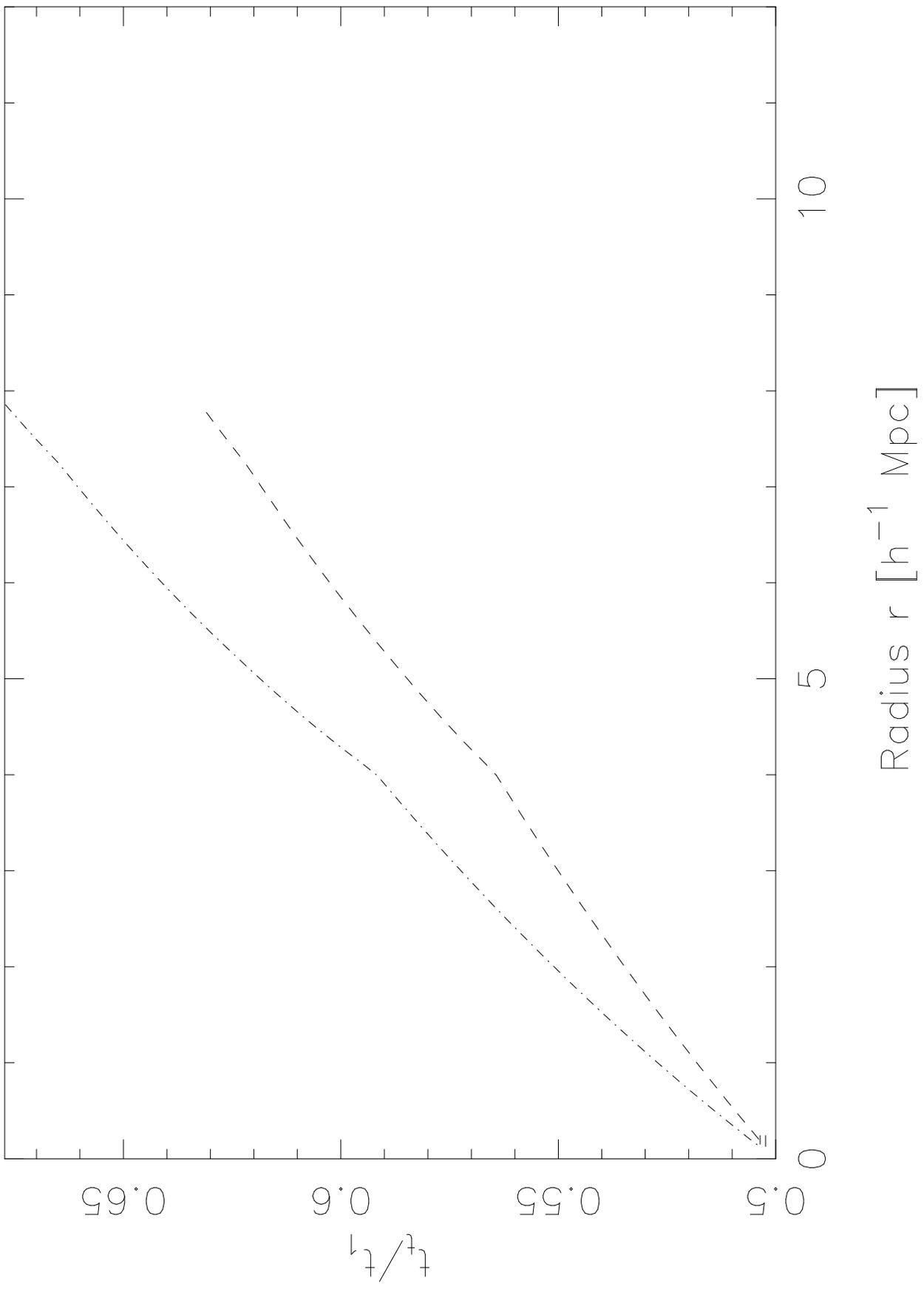
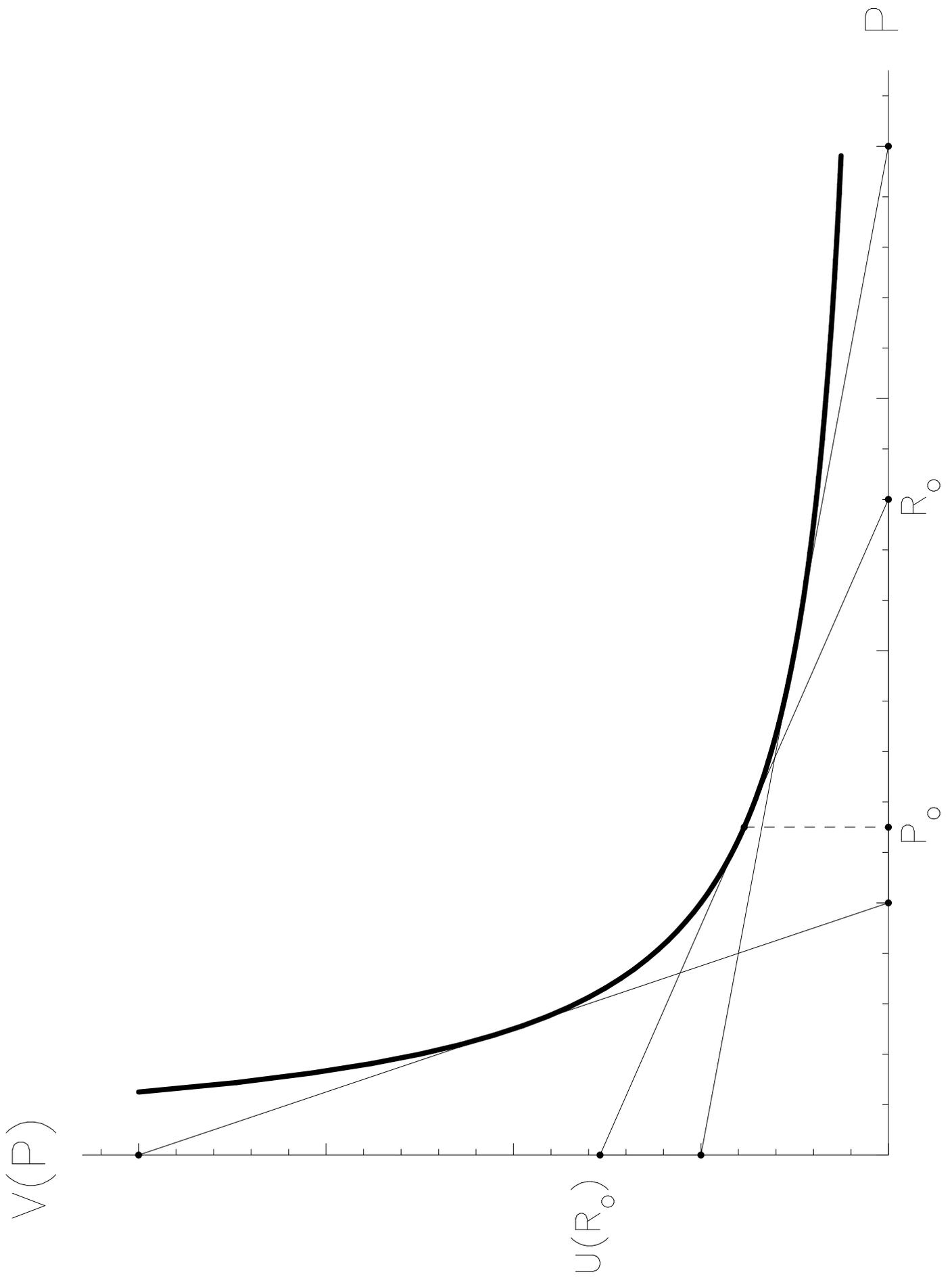
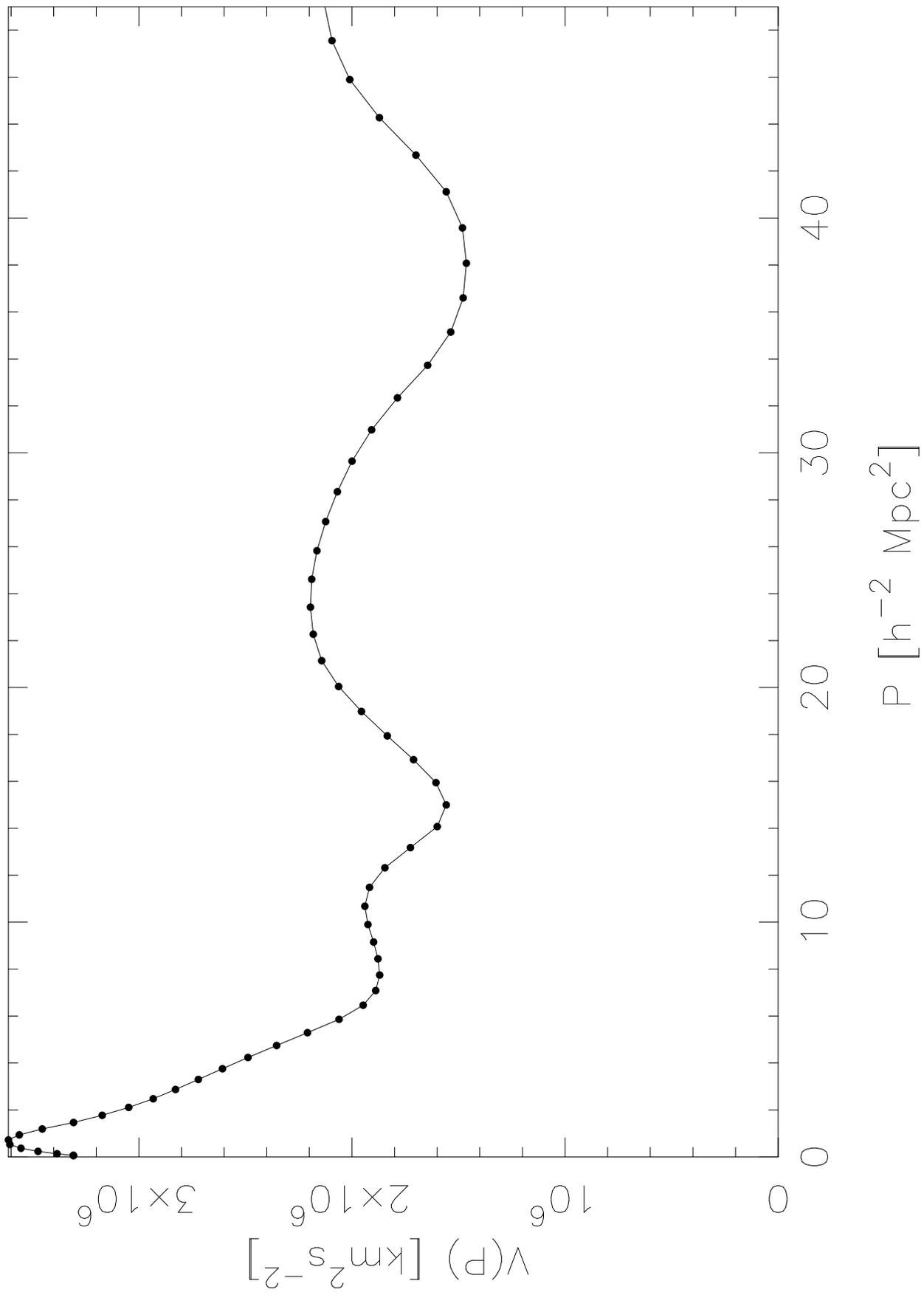
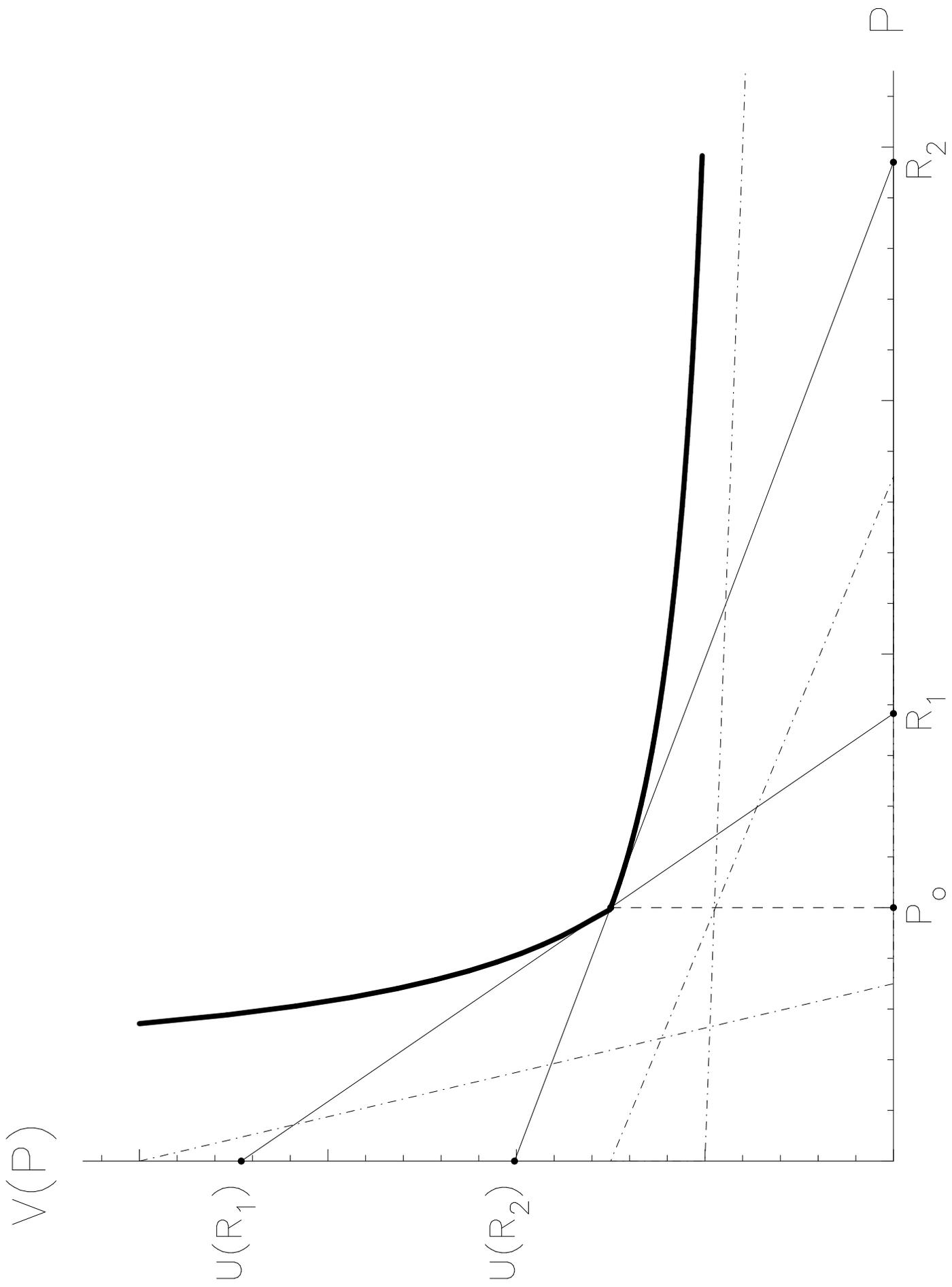
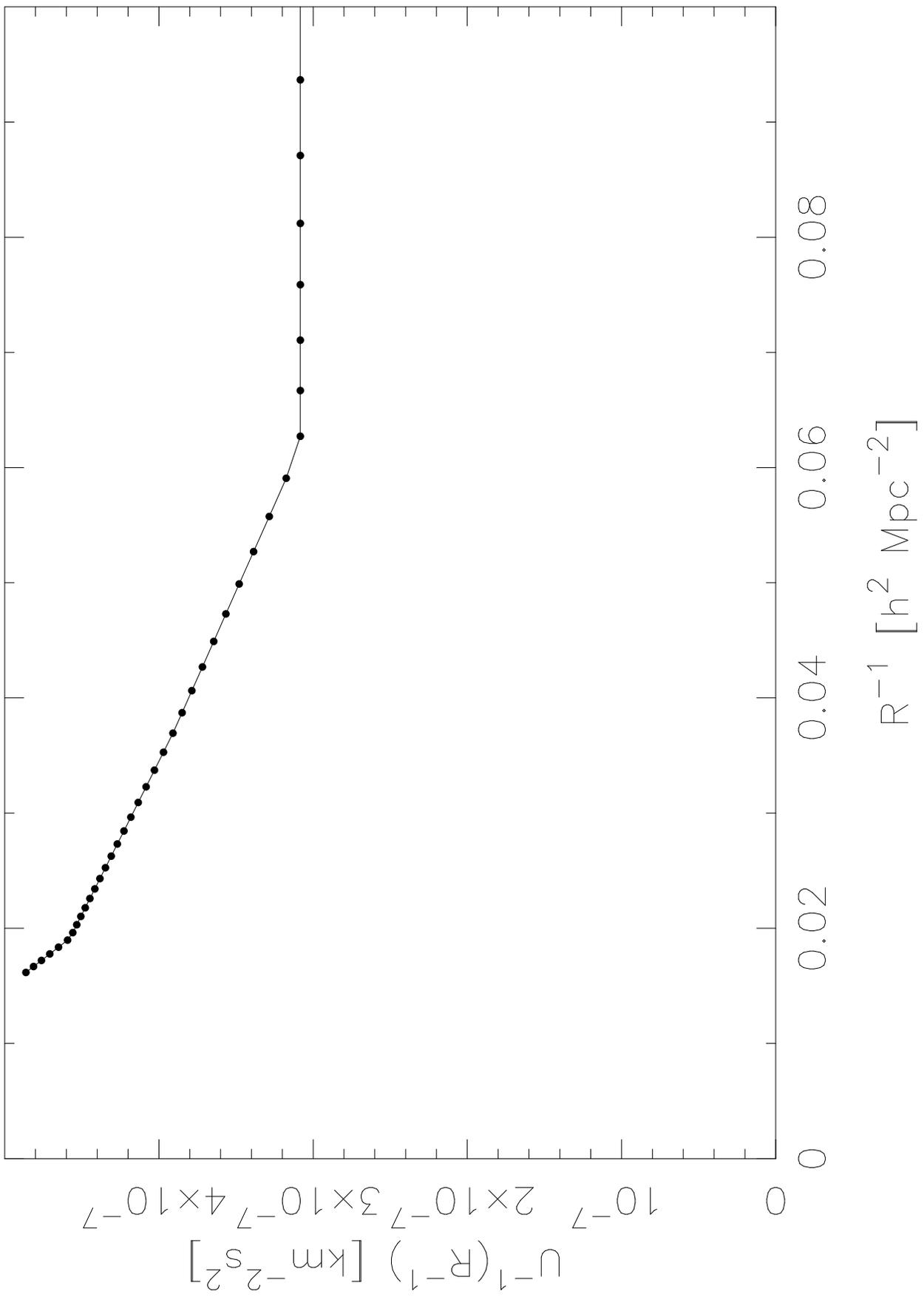
\end{document}